# The Photon-Assisted Cascaded Electron Multiplier: a Concept for Potential Avalanche-Ion Blocking

J.F.C.A. Veloso[a], F.D. Amaro[b], J.M.F. dos Santos[b], A. Breskin[c], A. Lyashenko[c] and R. Chechik[c]

[a]*Physics Dept., University of Aveiro, 3810-193 Aveiro, Portugal*
[b]*Physics Dept., University of Coimbra, 3004-516 Coimbra, Portugal*
[c]*Dept. of Particle Physics, The Weizmann Institute of Science, 76100 Rehovot, Israel*

*Abstract*

We present a Photon-Assisted Cascaded Electron Multipliers (PACEM) which has a potential for ion back-flow blocking in gaseous radiation detectors: the avalanche from a first multiplication stage propagates to the successive one via its photons, which in turn induce photoelectron emission from a photocathode deposited on the second multiplier stage; the multiplication process may further continue via electron-avalanche propagation. The photon-mediated stage allows, by a proper choice of geometry and fields, complete blocking of the ion back-flow into the first element; thus, only ions from the latter will flow back to the drift region. The PACEM concept was validated in a double-MHSP (Micro-Hole & Strip Plate) cascaded multiplier operated in xenon, where the intermediate scintillation stage provided optical gain of ~60. The double-MHSP detector had a total gain above $10^4$ and energy resolution of 18% FWHM for 5.9 keV x-rays.

*Submitted to **JINST** – Journal of Instrumentation*

# 1. Introduction

In this work, we present the concept Photon-Assisted Cascaded Electron Multipliers (PACEM) that has a potential for blocking avalanche-induced ions in cascaded gaseous electron multipliers: the avalanche originating in a first element propagates to the following ones solely via its photons, which in turn induce photoelectrons in the next multiplication stage. By a proper choice of the gas, geometry and electric fields the photon-assisted propagation can be made very efficient, while the charge transport between the stages is completely blocked. Consequently only ions from the first multiplication stage will flow back to the drift volume (or to the photocathode, in case of a gaseous photomultiplier).

Light emitted from excited gas molecules, and in electron avalanches, has been used for more than two decades for signal amplification and/or readout in radiation detectors [1-7]. The optical recording of avalanche signals has the advantage of being mechanically and electrically decoupled from the detector and thus immune to high-voltage problems and insensitive to electronic noise. The pursue of suitable gas mixtures to be used with adequate photon readout systems, in various applications including gaseous photon detectors, has been for many years a major subject of research; the goal has been high light yield in the near-UV and visible ranges, compatible with the practical range of optical components and photon detectors [2-7]. Already in the late seventies several authors discussed the idea, reviewed in [2], of recording gas-scintillation photons (mainly from xenon) with another gaseous wire chamber filled with photosensitive vapours; the idea was to replace standard photomultipliers by large-area economic devices. In a recent work, the authors suggested a "hybrid photo-detector" for the visible light: gas scintillation in xenon, induced by photoelectrons emitted from a bialkali photocathode, was detected in a wire-chamber incorporating a CsI photocathode [8]. The idea was to prolong the bialkali photocathode life-time by its decoupling from the electron avalanche. In addition to their beneficial role in optically-recorded detectors, photon-mediated gas-avalanche build-up is one of the basic mechanisms in detectors operating in Geiger- and Geiger-like modes [9].

On the other hand, photon emission, particularly in the UV spectral range, is rather disturbing in most gaseous detectors; their interaction with the gas and with detector elements causes secondary, photon-feedback effects, usually limiting the detector's gain and performance. This drawback was overcome with the introduction of the Gas Electron Multiplier (GEM) [10]; the avalanche confinement within the GEM holes effectively hinders photon-mediated secondary processes, particularly in cascaded configurations [11,12], allowing for high gains; these were reached in cascaded GEM and Micro Hole & Strip plate (MHSP) [13] also in highly UV-scintillating gases, such as noble-gases [14] and $CF_4$ [15]. Intensive investigations of such multipliers, in these gases and often in combination with CsI photocathodes, have been carried out in the fields of gaseous photomultipliers [16, 17], neutron detectors [18], x-ray imaging devices [14, 19] and in dual-phase noble-liquid detectors [20].

The performances of gaseous detectors are also severely limited by avalanche-induced ions. There have been intensive efforts in recent years attempting to reduce to minimum the ion-backflow; numerous ideas and methods were investigated for blocking the ions during their drift process, by deviating them electrostatically from their original trajectory [21]; the ion-backflow fraction (IBF, the fraction of final-avalanche ions that flow back to the drift volume) could be reduced at best to $10^{-3}$ and $10^{-4}$, in DC and in pulse-gate modes, respectively. The best ion-blocking performances so far were reached in cascaded GEMs and MHSPs [21] and in Micromegas multipliers [22]. However, these IBF values are still too high for some applications, e.g. in Time Projection Chambers (TPC) and in gaseous photomultipliers, particularly for the visible-spectral range, as discussed in [21].

In this work we present an approach that totally blocks the ion back-flow to the first multiplying element, with only the ions from this stage flowing back to the drift field. We describe the method, provide the results of our preliminary studies and discuss the method's limitations, applicability and future research plans.

## 2. The Photon-Assisted Cascaded Electron Multiplier

The Photon-Assisted Cascaded Electron Multiplier (PACEM) concept is shown in Fig. 1, e.g. for a double-element MHSP multiplier investigated within this work. In this scheme, the two elements are optically coupled, within the same gas volume, through a charge-blocking mesh-electrode G1. Radiation-induced ionization electrons, deposited in the drift volume, are focussed into the holes of MHSP1 and create an avalanche within the holes; a second avalanche occurs at the anode strips, where all electrons are collected [13]. Avalanche-generated photons induce photoelectrons from the CsI photocathode deposited on the top face of MHSP2; these create the final avalanche within the MHSP2 holes and at its anode strips. As discussed in [23], depending on the induction field below MHSP2, 20-40% of the final-avalanche ions drift back through the MHSP2 holes; these should be blocked by the G1 mesh, which is polarized (here grounded) such that the fields on its both sides are reversed. Under these conditions, regardless of the total gain of the cascaded detector, only ions originating from the first element will flow back to the initial drift region, or to the top face of the first multiplier (e.g. a reflective photocathode in a gaseous photomultiplier). This opto-coupling technique may be therefore applied between the first two stages of GPMs and TPC readout elements as a mean for reducing the ion backflow, provided ways and conditions are found to reduce the first's stage IBF to a minimal acceptable value.

Moreover, with a proper design of the photon-mediated stage, the number of photoelectrons induced on the photocathode at the second stage may exceed the number of primary electrons in the drift region, resulting in what we define as *optical gain*. It should be noted that the photon-mediated multiplication concept, shown in Fig.1, can naturally be applied in many different cascaded-multiplier configurations. The first element can be a wire chamber, a parallel-grid multiplier, a GEM etc.; it should be preferably one which provides the maximal light yield for a given gas gain as to generate the minimal number of ions per event. The cascade may comprise elements of different types, of which the number will be chosen according to the desirable total gain. Some advanced patterned ion-deviating elements are discussed elsewhere [21, 24, 25].

### 3. Results and discussion

The study was carried out with the setup of Fig.1, using $^{55}$Fe 5.9 keV x-rays. The wire mesh (80 μm diameter stainless steel wires with 900 μm spacing) was placed at a distance of 2 mm from each multiplier's face; the respective lengths of the drift and the induction regions are 17 mm and 2 mm. The detector was placed in a stainless steel vessel, evacuated to a vacuum of $10^{-6}$ mbar and then filled with xenon at 1 bar, continuously purified by convection with non-evaporable getters (SAES St707); the latter were placed in a small volume connected to the detector's vessel and operated at temperatures ~150ºC. The mesh, the detector window and the induction plane were grounded; the drift field was kept at $E_D=120$ V/cm; the transfer fields ($E_{T1}$ and $E_{T2}$), on both sides of the G1 mesh, were kept reversed to each other, as shown in Fig.1, $E_{T2}$ being set close to zero to ensure good efficiency in the focussing of the photoelectrons into the holes [16] and $E_{T1}$ about -3kV/cm. The voltage difference between the anode and cathode strips ($\Delta V_{aci}$) of both MHSPs as well as the voltages across the holes ($V_{holei}$) were varied in order to set the multiplier's gain.

The *charge* produced on the anode strips of MHSP1 and the avalanche-induced *scintillation-photon yield*, both dictated by $V_{hole1}$ and $\Delta V_{ac1}$, are proportional to the number of primary electrons deposited in the gas (drift region) by the 5.9 keV x-rays. The VUV scintillation photons ($\lambda$~172 nm in Xe) impinging on the CsI-photocathode deposited on the top surface of MHSP2, induce the emission of photoelectrons (QE~30% at 170 nm [26]). The charge multiplication in MHSP2 is dictated by $V_{hole2}$ and $\Delta V_{ac2}$, yields a final charge at the anode strips - proportional to the charge and light produced at MHSP1. Typical pulse-height distributions of signals collected on the anode strips of both MHSPs are depicted in Fig.2; the one of MHSP2 is a result of the photon-assisted cascaded multiplication. Good signal-to-noise and peak-to valley levels were obtained; energy resolutions of ~18% FWHM were recorded on both electrodes, indicating that the optical coupling does not deteriorate the charge multiplication properties.

The optical coupling was investigated by varying the scintillation yield of MHSP1 (varying $\Delta V_{ac1}$ and $V_{hole1}$) and keeping MHSP2 potentials and gain constant ($V_{hole2} = 370$V and $\Delta V_{ac2} = 200$V). The charge gain of MHSP2 was determined by grounding the mesh G1 and all MHSP1 electrodes, and biasing the MHSP2 top electrode to get $E_{T2}=100$ V/cm (see direction in Fig.1); this permitted collecting the charge pulses induced only by x-ray interactions in the gap between G1 and MHSP2 top electrode. For the above biasing voltages, a gain of $G_2 = 350$ was obtained. The same charge amplification chain was used for recording all signals.

Fig.3 presents the charge gain $G_1$ obtained at MHSP1 anode strips and the total gain $G_{total}$ of the full, photon-assisted, cascaded multiplier, recorded at MHSP2 anode strips, as a function of $V_{hole1}$, (Fig.3a) and of $\Delta V_{ac1}$ (Fig.3b). $V_{hole1}$ and $\Delta V_{ac1}$ were kept in the safe, discharge-free range. Absolute gains above $10^3$ and $10^4$ were reached in MHSP1 and in the entire cascade, respectively.

The total number of electrons collected on the anode strips of MHSP2, $N_{total}$, can be written in terms of $G_{total}$ and the number of primary electrons $N_0$ produced in the 5.9 keV x-ray interaction, or in terms of the charge gain of MHSP2, $G_2$ ($G_2 = 350$ for the present conditions), and the number of photoelectrons $N_{pe}$ transferred through the holes:

$$N_{total} = N_0 \times G_{total} = N_{pe} \times G_2 \qquad (1).$$

The *optical gain*, $G_{opt}$, of the optical-coupling stage is defined as the ratio $N_{pe}/N_0$ and is given by $G_{total}/G_2$; it is depicted in both Figs.3a and 3b. *Optical-gain* values ~60 were recorded, demonstrating the feasibility and efficiency of the proposed light-assisted multiplier cascade method in xenon.

In Fig.3a the curves for $G_{total}$ and $G_1$ are parallel, indicating that the number of photons produced in the avalanche increases proportionally to the number of avalanche electrons collected on MHSP1 anode strips, within the investigated $V_{hole1}$ range. On the other hand, in Fig.3b the *optical gain* increase with $\Delta V_{ac1}$ is not proportional to that of $G_1$, namely the light production is not proportional to the charge multiplication at the strips. This can be explained by the increase of the reduced electric field (E/p) around the anode strips, which results in an increase of the Townsend coefficient, favouring electron-impact ionizations on photon-generating excitations of the gas molecules.

It should be noted that we used a MHSP as the first multiplier element, in the scheme shown in Fig.1, in order to induce scintillation at the anode strips. For comparison, we operated this first element also in a GEM-mode, namely with hole-multiplication only, keeping $\Delta V_{ac1}$=0. Fig.4 compares $G_{total}$ as a function of $V_{hole1}$, for $\Delta V_{ac1}$= 0 and 220V. For equal $V_{hole1}$ values the GEM-mode yielded a five-fold lower total gain compared to the MHSP-mode.

## 4. Conclusions

We demonstrated the operation principle of the Photon-Assisted Cascaded Electron Multiplier in Xe. Avalanche-induced photons from a first MHSP, impinging on a second, CsI-coated MHSP, yielded an optical gain of ~60 with a total double-stage gain of $10^4$. In this scheme, a properly biased intermediate grid blocked the transport of both electrons and back-drifting ions between the two multipliers.

This optical-gate concept could be used for reducing the ion backflow in cascaded electron multipliers. In such a scheme, only ions produced in the first multiplier will contribute to the ion backflow to the drift volume, or towards the photocathode in a gaseous photomultiplier. Methods are being investigated for reducing this residual IBF.

With the present configuration of double-MHSP in xenon we can easily estimate the residual IBF, following the experience gained with ion backflow measurements in multi-GEM and GEM/MHSP multipliers. From [23] we know that the ions produced in the anode-strip avalanche are partly collected on the neighbouring cathode strips and partly on the cathode mesh below the MHSP. Thus, for an induction field of -5kV/cm, a residual 20% fraction of ions (and up to 40% at -1kV/cm) will go back into the MHSP holes (see figure 6 in [23]). The fraction of ions escaping from the first-element holes depends on the hole voltage and on the drift field [27,28]; it was measured by us (Fig.5) for a single MHSP operated in GEM mode at 1atm. Ar/CH$_4$(5%). At $V_{hole}$=400V the IBF is 17% with $E_{drfit}$=0.3kV/cm and it varies almost linearly with $E_{drift}$. Thus for $E_{drift}$=1 or 0.1 kV/cm we should expect IBF~50% or ~5%, respectively.

At the highest operation gains in the present work, the first MHSP operates at a total gain of $3*10^3$, of which the multiplication in the holes and at the strips is about equal. This implies that the dominant contribution to the IBF is from the final avalanche at the strips, and we should expect ~400 or ~40 ions per event on the average, for $E_{drfit}$=1 or 0.1kV/cm, respectively. With a total detector gain of $2*10^4$ this result is equivalent to IBF=$2*10^{-2}$ or $2*10^{-3}$ at the corresponding drift fields. This result is, in fact, inferior to the best IBF values obtained so far by electrostatic DC means [25].

To further improve the IBF, ways for reducing the charge gain in the first multiplier should be found, but without loosing the primary-electron detection efficiency. Possible approaches should be modification of the present scheme by optimizing the geometry of the first MHSP, e.g. by enlarging/reducing the hole diameter. Another possibility is to reduce the gain on the strips, compromising the optical gain, but significantly reducing the number of ions back-flowing through the holes. In the present work for example, by reducing the multiplication on MHSP1 strips, its charge gain could be reduced (e.g. a factor 10) but with a smaller reduction in optical gain (e.g. only a factor 3) (see Fig 3b), thus improving the IBF by factor 10 without a significant effect on the energy resolution and other detector performances. In our most recent (yet unpublished) work, the fraction of ions escaping the hole could be further reduced by flipping the MHSP's strips towards the drift volume, which could be implemented in this PACEM concept. Other then hole-multipliers, different multiplier configurations could be studied for this purpose, as mentioned before, such as wire-chambers, parallel gaps, Micromegas, etc.; Micromegas thin-gap multipliers would have an advantage of electro-statically blocking the ions simply due to the favourable field ratio to the drift field [22]. However, such "open-geometry" multipliers may suffer photon-feedback effects. A wide parallel-gap multiplier operating as a scintillation gap may produce a favourable photon-to-ion yield; however, such an arrangement could spoil the localization resolution of the detector.

Most important for the successful operation of the Photon-Assisted Cascaded Electron Multiplier is to find adequate gas mixtures, having high scintillation yields – matching the spectral sensitivity of the photocathode. From our experience, CsI has the highest stability and largest quantum yield among practical photocathode candidates [26]. Xe, Kr and Ar and their mixtures have emission spectra matching the CsI response. $CF_4$, which is a good counting gas for hole-multipliers [15] emits only a fraction of its light in the CsI sensitivity range [29] and may therefore require detector's first-stage operation at higher gains.

Further investigations of this optical ion-blocking concept are in course; they mostly concentrate on the optimization of the gas mixtures and on ion blocking at the first multiplication stage.

**Acknowledgements**

This work was supported in part by ProjectPOCTI/FP/63441/2005 through FEDER and FCT (Lisbon) programs, by the MINERVA Foundation and by the Israel Science Foundation project 402/05. A. Breskin is the W.P. Reuther Professor of Research in peaceful use of atomic energy.

# Figures

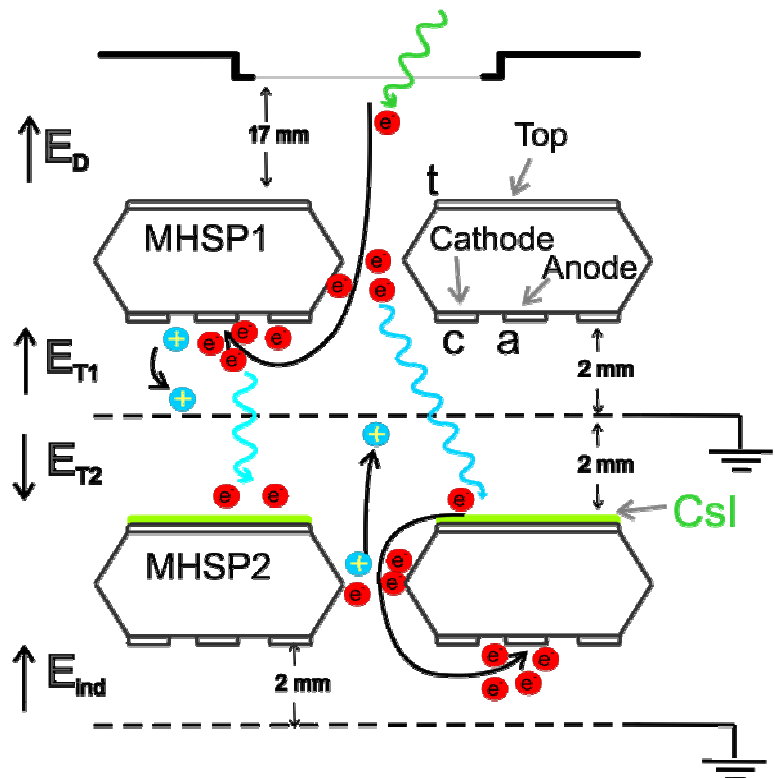

Fig.1 – Schematic layout of a double-MHSP detector operated in the photon-assisted cascaded electron-multiplier mode. The electron avalanche initiated in MHSP1 emits photons that induce photoelectrons from a photocathode (e.g. CsI) deposited on MHSP2; they are multiplied in the latter creating the final avalanche on its anode. Back-drifting ions are all collected at the grounded mesh, G1.

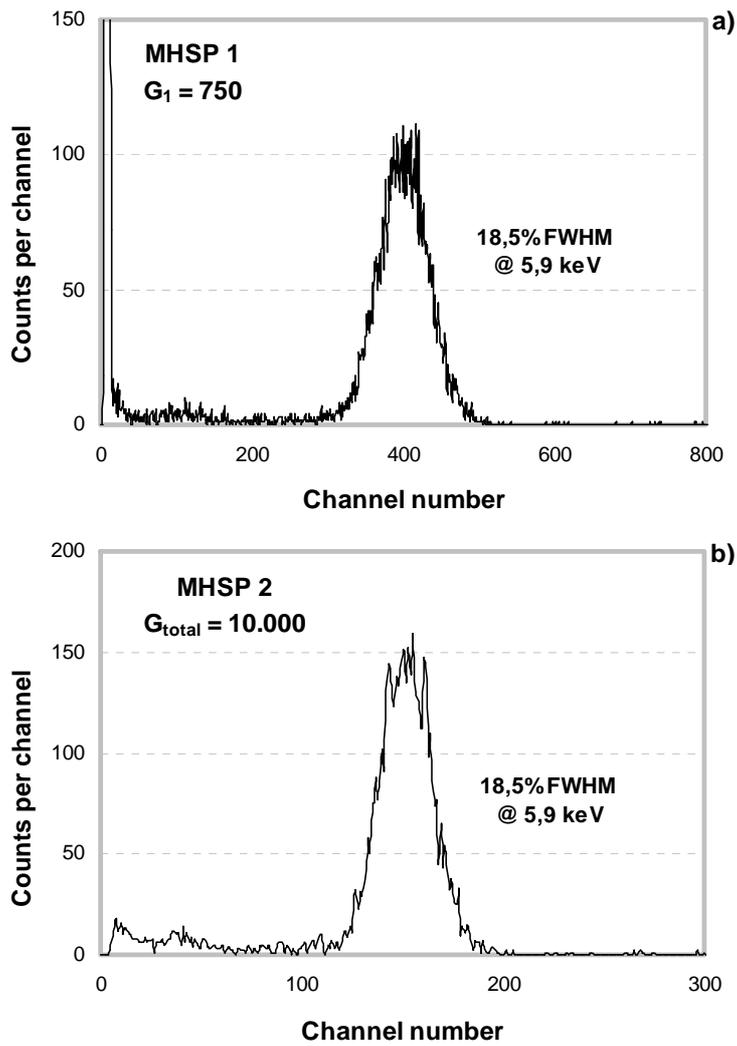

Fig.2 –Pulse-height distributions resulting from 5.9 keV x-ray interactions in the drift region of the cascaded detector of Fig.1, measured in 1atm Xe on the anode strips of MHSP1 (*a*) and MHSP2 (*b*).

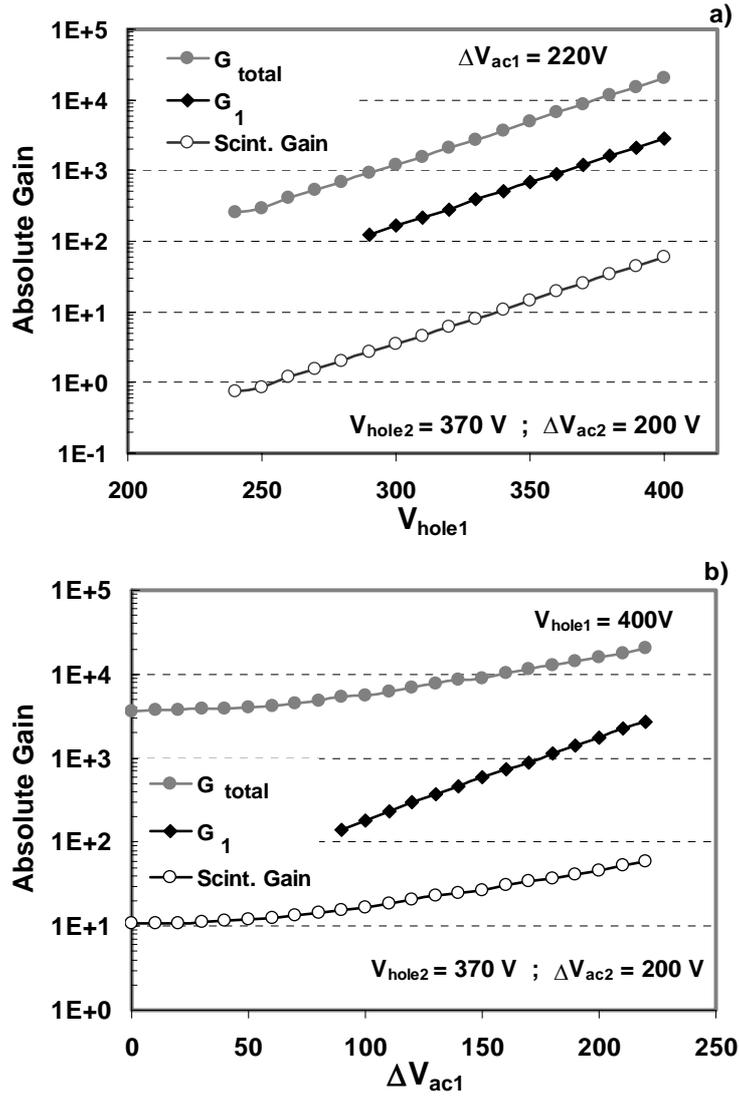

Fig.3 – Absolute gains obtained with 5.9 keV x-ray pulses recorded at MHSP1 ($G_1$) and MHSP2 ($G_{total}$) and the scintillation gain as a function of $V_{hole1}$ (a) and $\Delta V_{ac1}$ (b). The biasing voltages on MHSP2 were kept constant, $V_{hole2}$= 370V and $\Delta V_{ac2}$= 200V, corresponding to a measured absolute gain of $G_2$= 350.

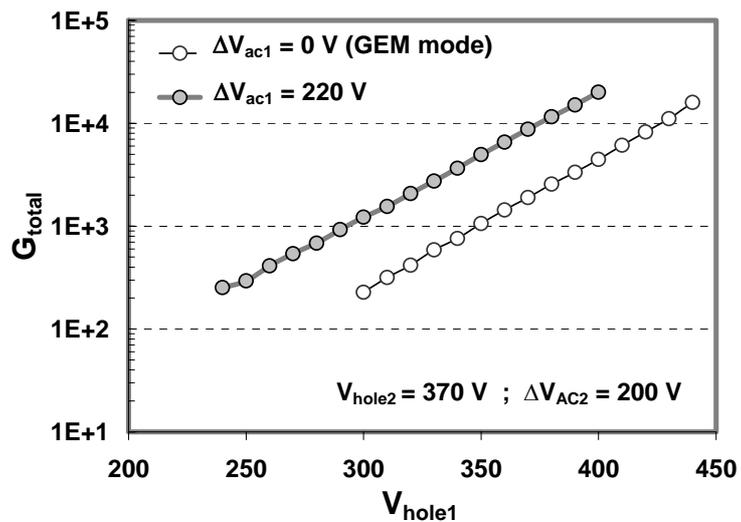

Fig.4 – Absolute Gain measured on the MHSP2 anode ($G_{total}$) as a function of $V_{hole1}$, for $\Delta V_{ac1} = 220$ V and for $\Delta V_{ac1} = 0$ V (MHSP operation in GEM mode) and for $V_{hole2}= 370$V and $\Delta V_{ac2}= 200$V.

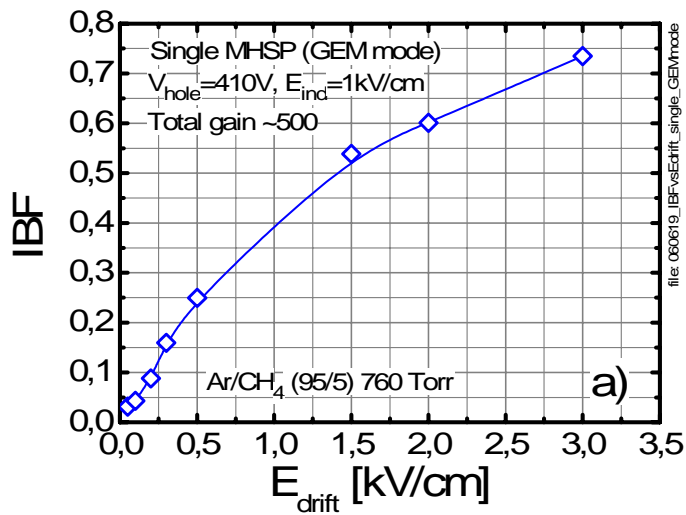

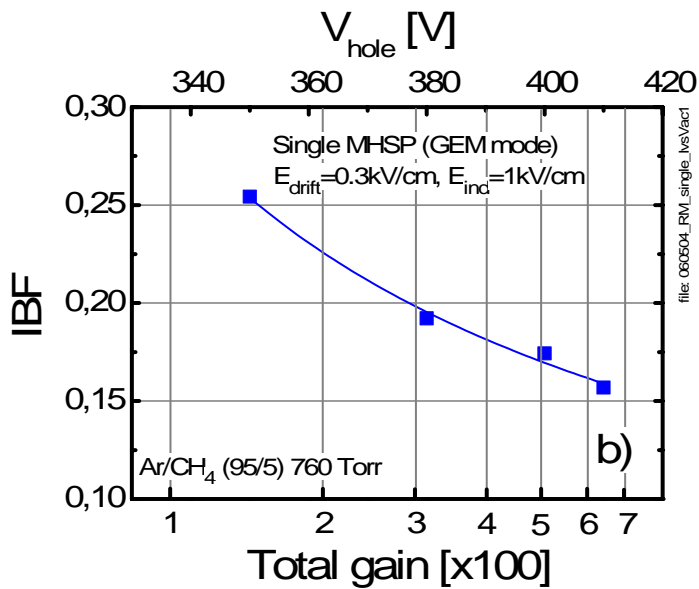

Fig.5 - Ion Back-flow Fraction (IBF) in a single MHSP element operated in GEM mode (no potential difference on the strips), in 1atm. Ar/CH$_4$(5%), as function of: a) the drift field; b) the hole gain (voltage). The IBF decreases almost linearly with V$_{hole}$ (increasing the dipole field) while the absolute number of ions increases exponentially (typical avalanche growth).